\documentclass[a4paper,10pt]{article}
\usepackage{mathrsfs}
\usepackage{epsfig}
\usepackage{amsmath}
\usepackage{amssymb}

\newcommand{\bfa}[1]{\mbox{\boldmath${#1}$}}

\newcommand{\bea}{\begin{eqnarray}}
\newcommand{\eea}{\end{eqnarray}}
\newcommand{\bnn}{\begin{eqnarray*}}
\newcommand{\enn}{\end{eqnarray*}}
\newcommand{\be}{\begin{equation}}
\newcommand{\ee}{\end{equation}}

\def\PACS{\par\leavevmode\hbox {\it PACS:\ }}%
\def\MSC{\par\leavevmode\hbox {\it MSC:\ }}%
\def\UK{\par\leavevmode\hbox {\it Keywords:\ }}%
\input xy
\xyoption {all}

\begin{document}

\title{Higher Spin Quaternion Waves in the Klein-Gordon Theory}

\author{S. Ulrych\\ Wehrenbachhalde 35, CH-8053 Z\"urich, Switzerland}
\date{September 9, 2012}
\maketitle

\begin{abstract}
Electromagnetic interactions are discussed in the context of the Klein-Gordon fermion equation.
The Mott scattering amplitude is derived in leading order perturbation theory 
and the result of the Dirac theory is reproduced except for an overall factor of sixteen. 
The discrepancy is not resolved as the study points into another direction.
The vertex structures involved in the scattering calculations indicate the relevance of a modified Klein-Gordon equation,
which takes into account the number of polarization states of the considered quantum field. 
In this equation the d'Alembertian 
is acting on quaternion-like plane waves, which can be generalized to representations of arbitrary spin.
The method provides the same relation between mass and spin that has been found previously by Majorana, Gelfand, and Yaglom
in infinite spin theories.

\end{abstract}

{\scriptsize\PACS{03.65.Pm;  11.80.-m; 12.20.Ds; 12.90.+b; 11.30.Ly}
\MSC{81R20; 81U99;  81V10; 81Q05; 81Q65}
\UK{Klein-Gordon equation; Mott scattering; Higher spin; Quaternionic quantum mechanics; Clifford algebra}}

\section{Introduction}
Quaternionic calculation methods are under permanent investigation and further development
to provide powerful mathematical tools for applications in various areas of physics.
Silberstein \cite{Sil12}, Lanczos \cite{Lan26}, Conway \cite{Con37}, G\"ursey \cite{Gur50, Gur58}, 
Finkelstein et. al. \cite{Fin62}, Edmonds \cite{Edm72}, Adler \cite{Adl80,Adl85,Adl95}, 
Horwitz and Biedenharn \cite{Hor84}, Baylis \cite{Bay99}, and many other authors inspired with their research on this subject other physicists. 
The number of publications concerning with quaternions or complexified quaternions is in fact endless. 
A bibliography with hundreds of entries has been compiled by Gsponer and Hurni in alphabetical \cite{Gsp08} and analytical form \cite{Gsp08b}.
Recent articles on quaternions in relativistic physics have been published by Demir et al. \cite{Dem10, Dem11}, Panicaud \cite{Pan11}, 
Bisht et al. \cite{Bis08, Bis10}, Christianto et al. \cite{Chr09}, and Tan{\i}\c{s}l{\i} \cite{Tan11}.

In the following text a quaternionic Klein-Gordon equation will be applied to fermions.
Some decades ago Feynman expressed his predilection for the Klein-Gordon fermion equation, because the representation of relativistic quantum mechanics with the Klein-Gordon equation by the method of path-integrals can be easily handled, 
but the Dirac representation is very hard to represent directly \cite{Fey48, Fey51, Fey58}.
The Klein-Gordon fermion equation, which is equivalent to the squared Dirac equation, has been derived by Kramers \cite{Kra33,Kra57} and Lanczos \cite{Lan33}. Field theoretic aspects of this equation have been investigated by Brown \cite{Bro58} and Tonin \cite{Ton59}.
See also Pilkuhn \cite{Pil03} for a detailed historic overview.
More examples of second order differential fermion equations can be found in
Biedenharn et al. \cite{Bie72}, Volkovyskii \cite{Vol71}, or recent articles published by \'Angeles and Napsuciale \cite{Ang11},
Nottale and C\'el\'erier \cite{Not09,Cel10}, and Delgado-Acosta and Napsuciale \cite{Del11}.

The Klein-Gordon method in \cite{Ulr05Fund} is
based on paravectors, which include the Pauli algebra as basis elements. Congruent complexified quaternion representations have been considered before also by Edmonds \cite{Edm72}, Baylis \cite{Bay80}, Sobczyk \cite{Sob81}, and
Sachs \cite{Sac82, Sac10}.
Recently, the Pauli algebra has been applied to fermion spin by Baylis et al. \cite{Bay10}, 
to polarization optics by Tudor \cite{Tud10}, and to semiconductor physics by Dargys, see \cite{Dar11} and references therein.

Quaternions are used as building blocks in the representation theory of Clifford algebras.
Starting point for the application of Clifford algebras in modern physics was the book of Hestenes \cite{Hes66}. For more details on further progress in this area see the recent articles of Hestenes \cite{Hes11} and Sobczyk \cite{Sob10}.
More information on Clifford algebras in physics can be found in Doran and Lasenby \cite{Dor03}, 
Rodrigues and Capelas de Oliveira \cite{Rod07}, Girard \cite{Gir07}, or Perwass \cite{Per09}.

In \cite{Ulr05Fund} quaternions have been complexified with the product of the complex and the hyperbolic unit to obtain the hyperbolic Pauli algebra. 
The hyperbolic numbers appear in fact in two representations. The complex number like representation has been considered by Shervatov \cite{She63}, Yaglom \cite{Yag68,Yag79}, Ryan \cite{Rya82}, Hucks \cite{Huc93}, Gal \cite{Gal02}, Yamaleev \cite{Yam07, Yam07b}, Catoni et al. \cite{Cat08,Cat11}, and many other authors.  On the other hand there is the
null plane representation in terms of the double field. An overview of the double field and its role 
in the representation theory of Clifford algebras has been provided by Porteous \cite{Por95}.
The hyperbolic numbers have been used by Khrennikov \cite{Khr10} and Nyman \cite{Nym11} to represent probabilistic data by quantum-like hyperbolic amplitudes. This research area has been detailed in \cite{Khr09,KhrB10}. The work on the Erlangen program of Kisil considers complex, parabolic and hyperbolic numbers and their geometric properties, see for example \cite{Kis07,Kis10,Kis11}.
Solutions of the Klein-Gordon equation with hyperbolic numbers have been studied by Bracken et al. \cite{Bra03} and 
Kravchenko et al. \cite{Kra08,Kra09}.

The following article applies the Klein-Gordon fermion theory as presented in \cite{Ulr05Fund, Ulr10} to calculate the Mott scattering amplitude. 
The result of the Dirac theory is reproduced except for a factor of sixteen. 
Though it might be possible to close the gap with a more detailed analysis, another method is proposed in
Section \ref{kleing}, which is based on the structure of the interaction vertex.
The Klein-Gordon equation is modified to include the number of possible polarization states of the considered quantum field. 
Quaternionic plane waves are introduced in Section \ref{polwave} as solutions of the equation, 
which could be denoted as Klein-Gordon equation for polarizable fields. 

The mathematical representation will be adjusted in Section \ref{notation} to align the
formalism more closely with the Clifford algebra approach. Because the article should serve as a bridge between the method specified in \cite{Ulr05Fund}
and the modified concepts, the notation is adjusted in the middle of the text and not from the beginning.

\section{Hyperbolic Pauli Algebra}
\label{alg}
The following considerations are based on the hyperbolic Pauli algebra,  
which is used to represent relativistic space-time vectors. A relativistic
vector $x$ is expanded into
\be
´x=e_\mu x^\mu .
\ee
The basis elements are defined as a
paravector algebra, which means one basis element of the algebra is the identity element.
The other basis elements are constructed of the Pauli algebra multiplied
by the hyperbolic unit $j\equiv \sqrt{+1}$
\be
\label{basis2}
e_\mu=(e_0,e_k)=(1, j\sigma_k)\;. 
\ee
The hyperbolic Pauli algebra $e_k$ is isomorphic to the Pauli algebra $\sigma_k$, but
includes the hyperbolic unit as an extra factor. The multiplication rules of the hyperbolic Pauli algebra are
\be
\label{algebra}
e_ke_l=\delta_{kl}1+ij\,\varepsilon_{klm}e^m\;.
\ee
In the same way as the Pauli algebra can be understood as the complexification
of the quaternions by the complex unit $i$,
the hyperbolic Pauli algebra can be understood as a complexification of
the quaternions by the factor $ij$.

The hyperbolic Pauli algebra corresponds to the real Clifford algebra
$\mathbb{R}_{3,0}$. 
The advantage of the hyperbolic Pauli algebra, compared to the standard Pauli algebra, is that the Clifford anti-involutions conjugation and reversion and the grade involution can be defined in a convenient form, when using a $2\times 2$ matrix representation of the algebra. Conjugation as matrix transposition 
and change of sign of the hyperbolic and complex units, reversion as matrix transposition and change of sign of the complex unit, 
and the grade involution as change of sign of the hyperbolic unit \cite{Ulr10}. 

In the mentioned $2\times 2$ matrix representation of $\mathbb{R}_{3,0}$ the scalar product can be defined 
with the conjugation anti-involution as
\be
\label{scalar}
x\cdot y = \frac{1}{2}Tr(\bar{x}y)\;.
\ee
Though the scalar product within Clifford algebras is normally defined in a representation free form,
the trace definition will be used to get a deeper understanding of an expression that appears in the scattering calculations of the following sections.

\section{Field equations and interactions}

The Lagrangian for free
noninteracting spinor fields has been defined in \cite{Ulr10} as
\be
\label{lagrange}
\mathcal{L}=\bar{\psi}(M^2-m^2)\psi\;.
\ee
The mass operator is represented in terms of the relativistic paravector algebra introduced in Section~\ref{alg}
\be
\label{massop}
M^2=p\bar{p}\;.
\ee
The coefficients of the momentum operator are $p_\mu=i\partial_\mu$.
They are invariant under conjugation
\be
\bar{p}_\mu=p_\mu\;.
\ee
The equation of motion, which is equivalent to the Klein-Gordon equation, can be derived from the above Lagrangian according to
\be
\label{mass}
M^2\psi=m^2\psi\;.
\ee
Interactions can be introduced as usual with the minimal substitution. The covariant derivative is defined as
\be
\label{subs}
P=p+A\;,
\ee
which ensures the invariance under gauge transformations.
A coupling constant has been omitted for simplicity.
The Lagrangian in  Eq.~(\ref{lagrange})
remains formally invariant after the minimal substitution. However, the mass operator
\be
\label{intermass}
M^2=P\bar{P}\;
\ee
is now equivalent to the quadratic Dirac operator with interaction terms \cite{Ulr05Fund}.

The field equation of the  gauge boson can be represented with the mass operator as given in Eq.~(\ref{massop}), which is now acting on a vector field
\be
M^2A=-J\;.
\ee
This representation includes all Maxwell equations \cite{Bay99,Ulr05Fund}.
The current $J$ is the source term of the interacting field. 
Contracted with a polarization vector $\varepsilon$ the current can be written as
\be
\varepsilon \cdot J=\bar{\psi}(p+A)\bar{\varepsilon}\psi-\bar{\psi}\varepsilon(\tilde{\bar{p}}-\bar{A})\psi\;.
\ee
Note that the momentum in the second term on the right side of the equation is acting to the left.
Using the plane wave expansion of $\psi$ defined in \cite{Ulr05Fund}, the matrix elements of the current 
can be deduced in leading order perturbation theory as
\be
\label{neuelec}
\langle f \vert \,\varepsilon \cdot J\,\vert i\rangle =\bar{u}_f(p_i\bar{\varepsilon}+\varepsilon\bar{p}_f)u_i\;.
\ee
The applied Feynman rules remind of scalar electrodynamics, but they include spin matrices and spinors.

\section{The scattering amplitude}
\label{scatter}
In this section the scattering amplitude is defined and further investigated in the context of unpolarized particle beams.
The scattering amplitude may be introduced in the following form
\be
{\cal M}=D_A(q)\sum_\varepsilon (\varepsilon\cdot J_1)(\bar{\varepsilon}\cdot\bar{J}_2  )\;,
\ee
with the sum over the possible polarizations of the exchanged gauge boson. The indices indicate vertices 1 and 2.
The propagator function of the gauge boson is denoted as $D_A(q)$.
The absolute square of the scattering amplitude may be written as
\be
|{\cal M}|^2 = |D_A(q)|^2\sum_{\varepsilon, \varsigma} {\cal N}_1 {\cal N}_2\;,
\ee
where the contributions arising from the two currents have the structure
\be
\label{namp}
{\cal N} =  (\varepsilon\cdot J)(\bar{\varsigma}\cdot\bar{J}) \;.
\ee
The indices have been dropped for clarity.
If the spin polarization of the scattered particles is not considered, one has to average the spin contributions of
the initial states and take the sum of the outgoing states. 
To calculate the squared scattering amplitude in first order perturbation theory without exchange term
the current as given in Eq.~(\ref{neuelec}) is inserted
\be
\langle {\cal \tilde{N}}\rangle=\frac{1}{2}\sum_{s_is_f}\bar{u}_i(p_f\bar{\varepsilon}+\varepsilon\bar{p}_i)u_f
\bar{u}_f(p_i\bar{\varsigma}+\varsigma\bar{p}_f)u_i\;.
\ee
The sum of the final spin states results in the identity element due to the completeness relation of the spinors \cite{Ulr05Fund}
\be
\langle {\cal \tilde{N}}\rangle=\frac{1}{2}\sum_{s_i} \bar{u}_i(p_f\bar{\varepsilon}+\varepsilon\bar{p}_i)(p_i\bar{\varsigma}+\varsigma\bar{p}_f)u_i\;.
\ee
For the initial states the Casimir trick can be used to simplify the expression according to
\be
\label{varn}
\langle {\cal \tilde{N}}\rangle=\frac{1}{2}Tr[(p_f\bar{\varepsilon}+\varepsilon\bar{p}_i)(p_i\bar{\varsigma}+\varsigma\bar{p}_f)]\;.
\ee 
The calculations have been performed so far based on the polarization vectors of the gauge boson to provide
a tensor free representation. 
This notation will not be applied further in the following text.

\section{Momentum dependence of the scattering amplitude}
In order to make the following calculations easily accessible, they are performed in a conventional tensor representation.
The squared absolute scattering amplitude is now written as
\be
\label{defscat}
|{\cal M}|^2 = |D_A(q)|^2 \,{\cal N}_{1\mu\nu} \,{\cal N}_2^{\mu\nu}\;,
\ee
where the completeness relation of the polarization vectors has been taken into account.
The involved tensors are defined as
\be
{\cal N}_{\mu\nu}=J_\mu \bar{J}_\nu\;.
\ee
The tensor free matrix elements of Eq.~(\ref{neuelec}) will be replaced by a current of the form
\be
\label{curelec}
\langle f \vert \, J_\mu\,\vert i \rangle =\bar{u}_f(p_i\bar{e}_\mu+e_\mu\bar{p}_f)u_i\;.
\ee

Before the calculations proceed, some useful relations are provided.
In analogy to the notation used by Doran and Lasenby \cite{Dor03} one may introduce the relationship
\be
\label{basis}
e_\mu\bar{e}_\nu=e_\mu\cdot e_\nu+e_\mu\wedge e_\nu\;.
\ee
The right side of this equation can be further transformed into
a conventional tensor representation, which is useful for explicit calculations.
The scalar product of the basis vectors is given as 
\be
\label{basissym}
e_\mu\cdot e_\nu=g_{\mu\nu}\;.
\ee
The wedge product is represented according to
\be
\label{basisasym}
e_\mu\wedge e_\nu=-i\sigma_{\mu\nu}\;.
\ee
The spin tensor, including the Pauli matrices $\sigma_i$, is calculated as \cite{Ulr05Fund}
\be
\label{sigspin}
\sigma_{\mu\nu}=
\left(\begin{array}{cccc}
\;0\;&\;-ij\sigma_1\;&\;-ij\sigma_2\;&\;-ij\sigma_3\;\\
\;ij\sigma_1\;&\;0\;&\;\sigma_3\;&\;-\sigma_2\;\\
\;ij\sigma_2\;&\;-\sigma_3\;&\;0\;&\;\sigma_1\;\\
\;ij\sigma_3\;&\;\sigma_2\;&\;-\sigma_1\;&\;0\;\\
\end{array}\right)\;.
\ee

With the help of this notation, the tensors involved in the scattering amplitude are calculated for unpolarized particle beams in analogy to Eq.~(\ref{varn}) 
\be
\label{tracescamp}
\langle {\cal \tilde{N}}\rangle_{\mu\nu}=\frac{1}{2}Tr[(d_\mu+iq^\rho\sigma_{\rho\mu})(d_\nu+i\sigma_{\nu\sigma}q^\sigma)]\;.
\ee
The spin sums have been evaluated using the Casimir trick and the completeness relation of the spinors.
Initial and final momenta have been replaced by the following definitions
\be
d=p_f+p_i\;,\hspace{1cm}q=p_f-p_i\;.
\ee
Because two terms in Eq.~(\ref{tracescamp}) vanish due to the trace operator,
the tensor can be represented in the form
\be
\label{tracescampres}
\langle {\cal \tilde{N}}\rangle_{\mu\nu}=d_\mu d_\nu - b_{\mu\nu}\;.
\ee
The second term on the right side of the equation is an abbreviation for
\be
\label{tensor}
b_{\mu\nu}= \frac{1}{2}q^\rho \,Tr\left(\sigma_{\rho\mu}\sigma_{\nu\sigma}\right) q^\sigma\;.
\ee
The trace has non-zero entries as displayed in Table~\ref{invo}.
\begin{table}
\begin{center}
\begin{tabular}{|c||c|c|c|c|c|c|}
\hline
1 & 0110 & 1001 & 0220 & 2002 & 0330 & 3003  \\
\hline
-1 & 1221 & 2112 & 2332 & 3223 & 1331 & 3113\\
\hline
ij & 0132 & 3201 & 0213 & 1302 & 0321 & 2103  \\
\hline
ij & 1023 & 2310 & 2031 & 3120 & 3012 & 1230\\
\hline
-ij & 0123 & 2301 & 0231 & 3102 &0312 &1203 \\
\hline
-ij & 1032 & 3210 & 2013 & 1320  &3021 & 2130 \\
\hline
\end{tabular}
\end{center}
\caption{Non-zero entries of $Tr\left(\sigma_{\rho\mu}\sigma_{\nu\sigma}\right)/2$. The first column shows the non-zero value, the other columns the indices, which have this value  \label{invo}}
\end{table}
The spin contributions thus provide the following structure
\be
 \frac{1}{2}Tr\left(\sigma_{\rho\mu}\sigma_{\nu\sigma}\right)=\eta_{\rho\mu\nu\sigma}+ij\epsilon_{\mu\nu\rho\sigma}\;.
\ee
The non-zero entries of the tensor $\eta_{\mu\nu\rho\sigma}$ correspond to the first two rows in Table~\ref{invo}. The second term is proportional to the anti-symmetric four-dimensional Levi-Civita symbol. 
Thus the tensor defined in Eq.~(\ref{tensor}) can be calculated as
\be
\label{rho}
b_{\mu\nu}= \eta_{\rho\mu\nu\sigma} q^\rho q^\sigma\;.
\ee
The antisymmetric contributions of the tensor vanish due to the symmetry of the two momentum vectors.

Now Eq.~(\ref{tracescampres}) can be inserted into Eq.~(\ref{defscat}) and the squared absolute scattering amplitude of the one-boson exchange Feynman diagram is calculated as
\be
\label{cross}
|\langle{\cal \tilde{M}}\rangle|^2 = |D_A(q)|^2\, [(d_1\cdot d_2)^2 - (d_1 d_1+d_2 d_2)\cdot b+b\cdot b]\;.
\ee
The dot product has been generalized to tensors of rank two. The notation
for the mixed terms is resolved according to
\be
\label{multdef}
d_id_i\cdot b=\eta_{\rho\mu\nu\sigma}q^\rho d^\mu_i d^\nu_i q^\sigma\;.
\ee
As a final remark it should be mentioned that the tensor $b$ can be written as the scalar product
\be
b_{\mu\nu}=(e_\mu\wedge q )\cdot( e_\nu \wedge q)
\ee
where Eqs.~(\ref{scalar}) and (\ref{basisasym}) can be used to resolve this notation.

\section{Mott scattering}
Based on the results of the previous section it is now possible to derive the squared absolute amplitude 
for an electron scattered of a much heavier particle, for example a proton. The proton with mass $M$ is at rest in the lab frame. 
The energy of the electron $E$ is related to the electron mass $m$.
\be
p_{1i}= \left(\begin{array}{c}E\\
0\\
0\\
p\\
\end{array}\right)\hspace{0.2cm}
p_{2i}= \left(\begin{array}{c}M\\
0\\
0\\
0\\
\end{array}\right)\hspace{0.2cm}
p_{1f}= \left(\begin{array}{c}E\\
0\\
p\sin{\theta}\\
p\cos{\theta}\\
\end{array}\right)\;.
\ee
Energy and momentum of the outgoing proton are determined by energy-momentum conservation.

With this parameterization one can calculate the vectors which contribute to the scattering amplitude of Eq.~(\ref{cross}).
Only terms proportional to the proton mass are considered in the following,
because the proton is very heavy compared to the electron.
Thus one finds
\be
(d_1\cdot d_2)^2\approx (4E M)^2=16 M^2(m^2 + p^2)\;.
\ee
The second and the last term of  Eq.~(\ref{cross}) can be neglected,
because these terms do not include any contributions proportional to the proton mass
\be
d_1d_1\cdot b \approx b \cdot b \approx 0\;.
\ee
In the third term
only the energy coordinate of $d_2$  contains the proton mass. With Eq.~(\ref{multdef})
and the $\eta$-tensor as given in the first two rows of Table~\ref{invo} one finds
\be
d_2d_2\cdot b\approx(2M)^2\bfa{q}^2=16M^2p^2\sin^2{\theta/2}\;.
\ee
Inserting now all contributions into Eq.~(\ref{cross}) the following result is obtained
\be
\label{poimat}
 |\langle{\cal \tilde{M}}\rangle|^2 = 16 |D_A(q)|^2M^2(m^2 + p^2\cos^2{\theta})\;.
\ee
The scattering amplitude needs to be multiplied by a phase space factor to derive the Mott formula.
However, the phase space calculation rules of the applied method have not been investigated so far.

\section{Dirac Theory}
The calculation of the Mott formula within the Dirac theory can be found in many textbooks, see for example Griffiths \cite{Gri87}.
The corresponding result for the scattering amplitude is
\be
 |\langle{\cal \tilde{M}}\rangle|^2 = |D_A(q)|^2\, M^2(m^2 + p^2\cos^2{\theta})\;.
\ee
The result differs by a factor of sixteen from the result of the Klein-Gordon fermion theory in Eq.~(\ref{poimat}).
The discrepancy could be resolved by a more in depth analysis of the Klein-Gordon approach.
Because the mass operator in Eq.~(\ref{intermass}) is equivalent to the squared Dirac operator both methods should provide the same cross sections.
However, even a solution of the factor problem is not able to invalidate the argument that the
Klein-Gordon fermion theory is only an alternative representation
of the Dirac theory and does not provide any new or deeper insights into physics.  

The rest of this paper thus considers a modified method, which is motivated by the mathematical structures
that appeared in the scattering calculations of the previous sections.
These calculations further indicate that the method specified in \cite{Ulr05Fund} should be aligned closer with 
other Clifford algebraic and quaternionic approaches, see for example Adler \cite{Adl95} or Baylis \cite{Bay99}.

\section{Alignment with Clifford algebraic methods}
\label{notation}
In the theory of Clifford algebras the pseudoscalar is obtained by multiplication of all available basis elements.
The pseudoscalar may be introduced within the hyperbolic Pauli algebra representation of $\mathbb{R}_{3,0}$ as
\be
\imath=e_1\bar{e}_2 e_3\;.
\ee
Using the basis elements as defined in Section \ref{alg} one finds
\be
\imath= -ij\;.
\ee
Equation~(\ref{algebra}) can be rewritten in terms of the pseudoscalar
\be
\label{algebranew}
e_k\bar{e}_l=-\delta_{kl}1 + \imath\varepsilon_{klm}e^m\;.
\ee
This equation indicates how the hyperbolic Pauli algebra is applied in its relativistic context.
The next modification is the redefinition of the tensor in Eq.~(\ref{sigspin})
\be
\label{newsigspin}
\sigma_{\mu\nu}=
\left(\begin{array}{cccc}
\;0\;&\;\imath e_1\;&\;\imath e_2\;&\;\imath e_3\;\\
\;-\imath e_1\;&\;0\;&\;e_3\;&\;-e_2\;\\
\;-\imath e_2\;&\;-e_3\;&\;0\;&\;e_1\;\\
\;-\imath e_3\;&\;e_2\;&\;-e_1\;&\;0\;\\
\end{array}\right)\;.
\ee
The elements of the Pauli algebra have been replaced in the spin tensor by $e_i= j\sigma_i$. 
The definition of the wedge product for the basis vectors needs to be adjusted accordingly.
\be
e_\mu\wedge e_\nu=\imath\sigma_{\mu\nu}\;.
\ee
The generators of the spin angular momentum are introduced as in \cite{Ulr05Fund}
\be
\label{polten}
s_{\mu\nu}=\frac{\sigma_{\mu\nu}}{2}\;,
\ee
but now based on the tensor $\sigma_{\mu\nu}$ as defined in Eq.~(\ref{newsigspin}). 
Within this representation the Pauli-Lubanski vector may be defined as
\be
\label{lubadef}
w_\mu=\epsilon_{\mu\rho\sigma\nu}s^{\rho\sigma}p^\nu=
\imath\sigma_{\mu\nu}p^\nu=e_\mu\wedge p\;.
\ee
Note, that the Pauli-Lubanski vector is multiplied essentially by a factor of two 
compared to definitions in common textbooks like Itzykson and Zuber \cite{Itz06}.
The last expression in Eq.~(\ref{lubadef}) represents the coordinates of the Pauli-Lubanski vector as biparavectors,
which can be considered as planes  formed by the momentum and the basis vectors of the coordinate system. See Baylis \cite{Bay99} for more details
on the geometric interpretation of biparavectors.

\section{Klein-Gordon equation for polarizable fields}
\label{kleing}
The following considerations are based on the observation 
that the Pauli-Lubanski vector is hidden in the vertices of the previously investigated scattering problem.
One may have a look at the matrix elements of Eq.~(\ref{neuelec}). 
For example the second term on the right side of the equation gives rise to the following representation
\be
\varepsilon\bar{p}=\varepsilon\cdot(p +w)\;,
\ee
which can be derived with Eq.~(\ref{basis}) and the above definition of the Pauli-Lubanski vector $w$.
As the combination of momentum and Pauli-Lubanski vector appears in the current, one could ask whether this
structure plays a more important role in the momentum representation of the wave equation.
Thus instead of using the mass operator in its original form, one could think of an
operator defined as
\be
\label{lubamass}
{\cal O}=(p+w)\cdot(p+w)\;.
\ee
One has to keep in mind that in this notation a conjugation is hidden, which evaluates
according to 
\be
\bar{w}_\mu=-w_\mu\;.
\ee
Together with the orthogonality of Pauli-Lubanski vector and momentum the following straightforward calculation
is leading to
\be
\label{lubasquare}
{\cal O}=p^2-w^2=m^2+4m^2s(s+1)=(nm)^2\;,
\ee
with $n$ the number of possible polarization states of the considered quantum field
\be
\label{spinxpol}
n=2s+1\;.
\ee
Equation~(\ref{lubasquare}) is valid for the $n=2$ spin representation, but it can be generalized also to an arbitrary number of polarization states,
including $n=1$ for spinless particles.
The wave equation in a momentum space representation could have the compact form
\be
\label{field}
{\cal O}\psi=(nm)^2\psi\;.
\ee
This representation assigns the matrix structures in the wave equation to the Pauli-Lubanski vector.

To give a representation of the wave equation in position space, the d'Alembertian is introduced as
\be
\label{Alem}
\Box=\partial\cdot\partial\;.
\ee
The Klein-Gordon equation for polarizable fields is then defined accordingly
\be
\label{newbasic}
[\Box +(nm)^2] \psi=0\;.
\ee
For spinless particles the equation reduces to the ordinary Klein-Gordon equation.
The plane wave solution $\psi$ will be discussed in the following section.

One may compare the modified wave operator with the mass operator introduced  in Eq.~(\ref{massop}).
In the new notation one finds
\be
\label{oldmass}
M^2=p\cdot (p+w)\;.
\ee
Without interactions the spin structure collapses to $M^2=p^2$ due to the orthogonality of momentum and Pauli-Lubanski vector.
After interactions have been introduced one obtains the squared Dirac operator with interaction terms \cite{Ulr05Fund}.
Thus one may understand Eq.~(\ref{oldmass}) in a wider sense as a synonym for the Dirac method.
By comparison with Eq.~(\ref{lubamass}) the difference to the modified wave operator becomes obvious.

The difference between the approaches can be explained alternatively if one traces the generalized mass term back to the initial
ingredients and writes
\be
\label{Meng}
-\Box (E_4)= C_2(E_4)= p^2 - w^2 = C_2(T_4) + C_2(E_4/T_4)\;.
\ee
In this equation the quadratic Casimir operators with respect to the Poincar\'e group $E_4$, the translational group $T_4$, and the coset space $E_4/T_4$
are displayed. Equation~(\ref{Meng}) has
been introduced in analogy to the Laplacian used by Meng \cite{Wen03} in investigations 
of the Quantum Hall Effect in higher dimensions.
To make the analogy with Meng more obvious one may consider the d'Alembertian of the ordinary Klein-Gordon equation as a Laplacian
acting on the group of translations $T_4$ in Minkowski space-time
\be
\bigtriangleup(T_4) = -\Box (T_4)= C_2(E_4)-C_2(E_4/T_4)\;.
\ee
This equation has the structure as discussed by Meng \cite{Wen03}.
If one builds a mathematical method on top of a Laplacian in $T_4$, but the real
physical world in vacuum is $E_4$, there is the risk that additional mathematical structures
have to be introduced at a later stage to align the $T_4$ method with experimental observations.
This gives an argument to redevelop the methodology with the $E_4$ Laplacian 
of Eq.~(\ref{Meng}) from scratch.

\section{Quaternion waves}
\label{polwave}
Adler uses the notion of quaternion waves \cite{Adl86} which should be adopted in the following text
for the solutions of Eq.~(\ref{newbasic}). The term quaternion must be understood here in a generalized sense.
In fact it would be better to speak of quaternion-like plane waves.
They are introduced as
\be
\label{quaternionwave}
\psi=\exp{(-\imath x \bar{p})}\;.
\ee
This structure can be transformed into various mathematically equivalent representations.
A representation referring to the Clifford methodology can be derived with the help of Eq.~(\ref{basis})
\be
\psi=\exp{(-\imath(x \cdot p +l))}\;,
\ee
with the Clifford angular momentum \cite{Dor03}
\be
l = x \wedge p=\frac{1}{2}(x\bar{p}-p\bar{x})\;.
\ee
For the representation of spinless particles the relativistic angular momentum term can be discarded.

Another view on the plane wave is given by a tensor representation of the form
\be
\label{spinorbit}
\psi=\exp{(-\imath x_\mu p^\mu+s_{\mu\nu} l^{\mu\nu})}\;,
\ee
with the relativistic orbital angular momentum
\be
l^{\mu\nu}=x^\mu p^\nu - x^\nu p^\mu\;
\ee
and the spin angular momentum as defined in Eq.~(\ref{polten}).
Compared to the standard plane wave, which is related to translations only, the quaternion wave includes
translations, rotations, and boosts.
The representation which is most useful to reproduce the momentum space operator of Eq.~(\ref{lubamass}) is
\be
\psi=\exp{(-\imath x\cdot (p + w))}\;,
\ee
where the d'Alembertian of Eq.~(\ref{Alem}) is applied to the quaternion wave. 

The Pauli-Lubanski vector does not provide any new information once the momentum and the number of possible
polarization states of the considered quantum field are given. The proposed method thus describes
states of plane quaternion waves by the mass, the number of polarization states, and the momentum four-vector
\be
\label{staterepresent}
\vert\psi\rangle=\vert m,n,p\rangle\;.
\ee
The number of polarization states may be interpreted as the
intrinsic dimensionality of the quantum field. 
The quaternion waves refer to the Poincar\'e group as the gauge group of the method.
Spinors can be introduced in this context as elements of the vector bundle, which is associated to the Poincar\'e group principal bundle.

The exponent in Eq.~(\ref{quaternionwave}) can be transformed to a different overall sign, to the opposite order of momentum and position
vectors, or to the opposite location of the bar symbol. The second transformation changes the sign of 
the relativistic wedge product, the third transformation the
sign of a product of the form
\be
\label{newproduct}
x\wr p = \frac{1}{2}(x\bar{p}-\bar{x}p)\;.
\ee
The eight possible discrete states of the quaternion wave must be encoded in Eq.~(\ref{staterepresent}) when the 
mentioned three types of discrete transformations are considered. The transformations form the cyclic group 
$\mathbb{Z}_2\otimes\mathbb{Z}_2\otimes\mathbb{Z}_2$, which is isomorphic to the CPT group. 

\section{Higher spin}
\label{isospin}
The combination of Eq.~(\ref{spinxpol}) with the constant term in Eq.~(\ref{newbasic}) provides
the same relationship between spin and mass as initially found by Majorana \cite{Maj32}
and later also by Gelfand and Yaglom \cite{Gel48}
\be
\label{massenspin}
m=\kappa\left(s+\frac{1}{2}\right)^{-1}\;.
\ee
See here also the work of Fradkin \cite{Fra65}, Casalbuoni \cite{Cas06}, and Bekaert et al. \cite{Bek09}.
The constant $\kappa$ has been introduced as
\be
\kappa=\frac{nm}{2}\;.
\ee
Equation~(\ref{massenspin}) points towards an equivalence of the
Klein-Gordon equation for polarizable fields with the mentioned methods, which also refer to an arbitrary-component 
relativistic wave equation in comparison to the finite-component Dirac equation.

Relativistic wave equations for particles of arbitrary spin have been introduced also by Dirac \cite{Dir36}, Fierz, and Pauli \cite{Fie39a,Fie39}.
Higher spin quaternion waves have been investigated by Edmonds \cite{Edm76}.
Varlamov considers higher spin in combination with the CPT group \cite{Var11}.

\section{Summary}
The Klein-Gordon fermion theory in the mass operator representation has been applied to electron-proton scattering. 
It has been shown that the scattering amplitude
 calculated with this method is in agreement with the Dirac theory except for an overall factor of sixteen.
In principle, the results provided by the mass operator should be identical due to the equivalence of mass and squared Dirac operator.
Further investigations are necessary to explain this discrepancy.

The structure of the vertex, as provided by the analysis of the scattering problem, indicates the relevance 
of a modified Klein-Gordon equation for polarizable fields.
The mass is multiplied in this equation by the number of possible polarization states. The modified Klein-Gordon equation
cannot be aligned with a squared Dirac operator, but equivalence to concepts derived by Majorana, Gelfand, and Yaglom is indicated.

\section{Acknowledgements}
The author would like to thank Evi Bender, Ewald Lehmann, Andrei Khrennikov, and Vladimir V. Kisil for inspiring discussions and continuous support.

\end{document}